\documentclass[10pt]{iopart}

\usepackage{iopams}

\usepackage{graphicx}
\usepackage{xcolor}
\usepackage{todonotes}
\usepackage{subcaption}
\usepackage{longtable}

\newcommand{\lvec}{\mathbf{l}}
\newcommand{\svec}{\mathbf{s}}

\newcommand{\drm}{\mathrm{d}}

\begin{document}

\title[Prediction of lab-grown tissue properties using deep learning]{Rapid prediction of lab-grown tissue properties using deep learning}

\author{Allison E. Andrews, Hugh Dickinson and James P. Hague}

\address{School of Physical Sciences, The Open University, Milton Keynes, MK7 6AA}
\ead{Jim.Hague@open.ac.uk}
\vspace{10pt}
\begin{indented}
\item[]March 2023
\end{indented}

\begin{abstract}
The interactions between cells and the extracellular matrix are vital for the self-organisation of tissues. In this paper we present proof-of-concept to use machine learning tools to predict the role of this mechanobiology in the self-organisation of cell-laden hydrogels grown in tethered moulds. We develop a process for the automated generation of mould designs with and without key symmetries. We create a large training set with $N=6500$ cases by running detailed biophysical simulations of cell-matrix interactions using the contractile network dipole orientation (CONDOR) model for the self-organisation of cellular hydrogels within these moulds. These are used to train an implementation of the \texttt{pix2pix} deep learning model, reserving $740$ cases that were unseen in the training of the neural network for training and validation. Comparison between the predictions of the machine learning technique and the reserved predictions from the biophysical algorithm show that the machine learning algorithm makes excellent predictions. The machine learning algorithm is significantly faster than the biophysical method, opening the possibility of very high throughput rational design of moulds for pharmaceutical testing, regenerative medicine and fundamental studies of biology. Future extensions for scaffolds and 3D bioprinting will open additional applications.
\end{abstract}

%
%
%
%
%

\section{Introduction}

There is a need for fast, efficient algorithms to understand and predict the self-organisation resulting from interactions between cells and extracellular matrix (ECM), particularly in cultured tissues. The extracellular matrix provides structure to tissue and is important for tissue development, maintenance and repair \cite{kular2014a}. Cell-matrix interactions allow for the feedback of forces between the cells and the ECM leading to self-organisation \cite{Hayrapetyan2020}. This mechanobiology is a critical part of the behaviour of cells in tissues \cite{kular2014a}. Cultured tissues mimic the 3D structures in real tissues and have applications in regenerative medicine \cite{bajaj2014a}, testing in pharmacology \cite{weinhart2019a,jensen2018a}, cultured meats \cite{benarye2019a} and are useful for the study of fundamental biology. The goal of this paper is to show proof-of-concept for the use of machine learning techniques to make predictions regarding the shapes and self-organisation of artificial tissues grown in moulds with tethers. 

Such simulations may be useful for designing moulds and scaffolds suitable for tissue growth with specific characteristics, such as highly aligned cells. From an experimental point of view, it can be difficult and time consuming to design moulds and scaffolds to make cultured tissue with realistic characteristics. Several growth strategies for such tissue are available \cite{bajaj2014a}. Tethered cell-laden hydrogels can be a convenient way to make such tissues and have specific applications in high-throughput screening (e.g. muscle \cite{capel2019a}) and growing small volumes of tissue with potential regenerative medicine applications (e.g. nerves \cite{georgiou2013} and cornea \cite{mukhey2018}). Design of moulds and scaffolds for these applications by trial and error is time consuming due to the time it takes to manufacture the moulds, and then grow and analyse tissues grown within them. Biophysical models offer the possibility to speed up this process by testing the properties of moulds computationally to determine how tissue grows within specific mould designs.

The contractile-network-dipole-orientation model CONDOR is a microscopic model for describing the self-organisation resulting from cell-matrix interactions in tissues with polarized or elongated cells \cite{hague2019a}. The CONDOR model has a number of key features. The interaction between the cells and the ECM is dictated by the symmetries of the cells. It is microscopic, allowing it to be easily extended e.g. for higher order interaction symmetries and biological variation. It is relatively fast for a mathematical model. The self-organisation of large numbers of cells (currently up to around 100000) can be simulated allowing macroscopic tissue sizes to be approached. Validation by comparing with tethered cellular hydrogels (of glial cells) has shown that it accurately predicts how cell-matrix interactions drive the shaping and self-organisation of tissues. The CONDOR model typically takes between 24 hours and 1 week on a single CPU core to simulate up to 100000 cells. 

CONDOR simulations are complementary to widely available vertex models \cite{silvanus2017} and voronoi models \cite{bi2016}, which typically neglect ECM and are limited to $\sim 10^3$ cells. We also note the use of continuum models to predict the alignment of stress fibers in artificial tissue constructs \cite{deshpande2006,pathak2008,obbinkhuizer2014,legand2009}. Also available are force-dipole approaches with fixed cell positions that cannot predict tissue deformations \cite{schwarz2013}.

Machine learning techniques have the potential to speed up the process of predicting self-organisation in tissues to enable rapid rational design of moulds and scaffolds for cultured tissue. The use of Deep Learning models to efficiently and accurately approximate complex functions has been demonstrated in numerous scientific contexts including astronomy \cite{2021MNRAS.504.2603T,2022arXiv220613306D}, medicine \cite{frid2018gan, wolterink2017deep}, ecology \cite{2020RemS...12..901D,2020TCry...14..565B} and Earth observation \cite{2020NatSR..10.1322B,2018RSEnv.204..509B}.  In this work, we use CONDOR simulations to train an implementation of the \texttt{pix2pix} generative adversarial network architecture \cite{2016arXiv161107004I}. Given an input image comprising simple polygonal shapes that represent a mould and zero or more tethers, our trained \texttt{pix2pix} model is able to rapidly predict pixellated representations of the cell-laden hydrogels' equilibrium properties (i.e. the steady-state self-organisation of cells in the hydrogel) that accurately match CONDOR simulation output. 

To the best of our knowledge, in this paper we present the first application of deep learning to the prediction of self-organisation in tissues. Machine-learning techniques have already been widely adopted to streamline and accelerate many aspects of tissue engineering research see e.g. Ref. \cite{ten.tea.2022.0128} and references therein. These previous applications process large experimental datasets to \textit{infer} manufacturing techniques and engineered tissue parameters that are best suited for specific applications like regenerative medicine. Our method is fundamentally different because it does not rely on experimental datasets for inference. Instead, by  approximating the output of a proven biophysical model, we can accurately \textit{predict} the detailed structures and properties of tissues that arise from their self-organisation within different shaped moulds. Our detailed predictions can be used to compute the bulk tissue parameters that other methods infer, and thereby identify mould shapes that produce tissues with suitable properties for specific applications. The predictions of our model can also be compared with real experimental datasets to verify their realism and, consequently, their utility in real-world contexts. 

%
%
%
%
%
%
%

\section{Methods}

\subsection{Automated mould creation for training data}
\label{sec:mouldgeneration}

In order to train the machine learning algorithm we require a large quantity of data representing tissue growth in 1000s of moulds. To create these examples, we have developed a script to generate automated moulds; this script creates mould outlines and positions tethers in a quasi-random manner.

We generate mould shapes and tethering posts by randomly generating a set of 3 to 6 coordinates on a two-dimensional plane in the range $\left[0,1\right]$ for both axes. This sequence of coordinates are used to define a simple polygon or convex hull. We reject shapes which have a cross-sectional area with less than $0.5$ of the total canvas area to ensure they have reasonable dimensions, and subsequently regenerate coordinates. Following this, we generate 3 tethers at random positions within this polygon, each with a radius between $0.02$ and $0.065$ total canvas widths.


We then duplicate and transform the initial shape and tethers to produce a number of types of symmetry in the final mould. We use a total of 5 different duplication regimes; the first two involve reflections on either one or both axes to produce moulds which we label as single mirrored and double mirrored respectively. By duplicating with rotational transformations we can produce moulds with cyclic symmetry, or also dihedral symmetry if the initial mould is duplicated with a reflection beforehand. In all these cases, the origin for the mirror axes or rotations is a random position within the initial mould area. Finally we also use a set of random rotations each around different origin coordinates to produce an irregular and non-symmetric mould. Moulds produced by each regime are generated with equal probability. For moulds with either rotational symmetry, the order of symmetry can be between 2 and 6, again generated with equal probability.

\begin{figure}[h]
\centering
{
\includegraphics[width=\textwidth]{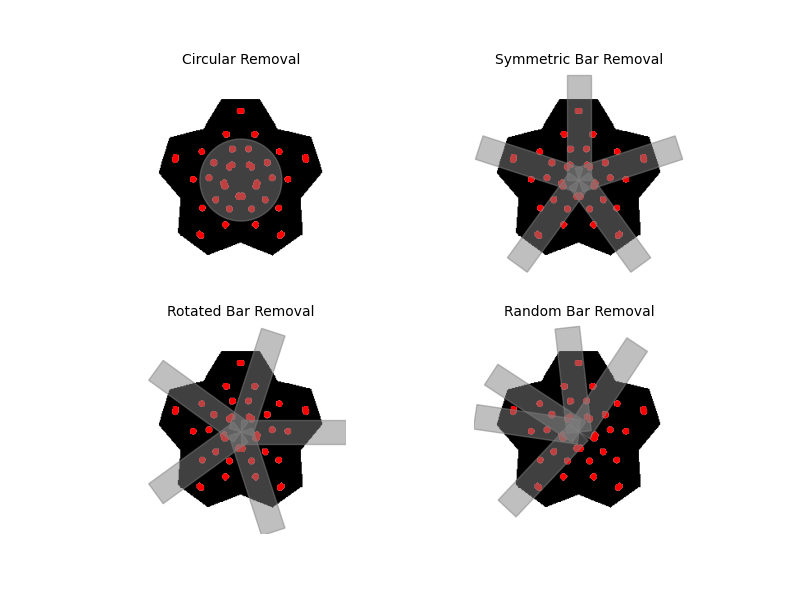}
\caption{Different tether removal methods used for mould generation.}
\label{fig:tether_removal_examples}
}
\end{figure}

Tethers are pruned after the duplication process to both reduce the total number present within the final mould, as well as to modify further the distribution and symmetry within the mould. This is achieved by defining an exclusion area for tethers for which we choose from several types. The first of these is a circular area placed in the centre of the mould with a diameter between 0.2 and 0.8 times the maximum mould width (determined randomly). The second is with bars placed along either the mirror axes or rotational vectors extending out from the centre, with width between 0.1 and 0.4 of the maximum mould width; these normally retain the same symmetry of the mould, but a rotational offset can also be applied to change the symmetry of tethers, i.e. a mould with dihedral symmetry can then have tethers with only rotational symmetry. Finally, we can use this same bar removal but with random alignment to produce moulds with asymmetric tether distributions. A diagram of these different types of tether removal is shown in Fig.~\ref{fig:tether_removal_examples}. The removal method selected is chosen at random with probabilities based on the symmetry assigned to the mould, shown in Table~\ref{table:tether_removal_probabilities}. For the majority of generated moulds we discard those which are subsequently left with less than 4 tethers; we have however included a subset of 100 moulds which have been generated with no tethers in our training data.


\begin{table}[h]
\centering
\caption{Probabilities of tether removal selection applied. The five symmetry types are selected with equal probability. For cyclic and dihedral rotations, the rotation order is selected with equal probability between $2$ and $6$ inclusive.}
\begin{tabular}{ c|c|c|c|c|c|c } 
Symmetry & Selection & \multicolumn{5}{c}{Removal type relative selection frequency } \\
& frequency & None & Circular & Symmetric bars & Rotated bars & Random Bars \\
\hline
Random rotation & $0.2$ & $0.05$ & $0.5$ & n/a & n/a & $0.45$ \\
Single mirror & $0.2$ & $0.05$ & $0.45$ & $0.45$ & $0.05$ & n/a \\
Double mirror & $0.2$ & $0.05$ & $0.45$ & $0.45$* & $0.05$ & n/a \\
Cyclic rotation & $0.2$ & $0.05$ & $0.45$ & n/a & $0.45$ & $0.05$ \\
Dihedral rotation & $0.2$ & $0.05$ & $0.45$ & $0.3$ & $0.15$ & $0.05$ \\
\hline
\multicolumn{7}{l}{*In double mirror case, this can be either horizontal axis, vertical axis, or both.} \\
\end{tabular}

\label{table:tether_removal_probabilities}
\end{table}

\begin{figure}[h]
\centering
{
\includegraphics[width=\textwidth]{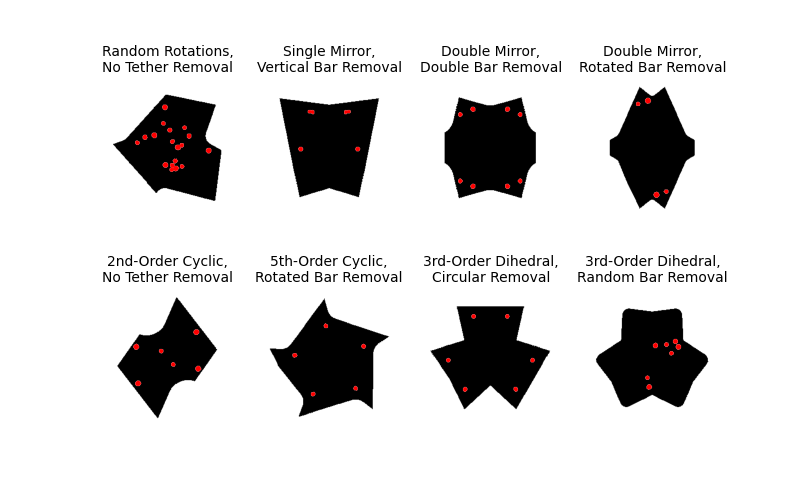}
\caption{A selection of mould and tether designs demonstrating the various symmetry and tether pruning regimes.}
\label{fig:final_mould_desing_examples}
}
\end{figure}

Finally, we round vertices in order to create curved edges in the mould design. Rounding of either convex and concave corners could be done independently, allowing for more variation in overall mould designs. Convex rounding was applied to 25\% of all moulds, while concave rounding was applied to 75\% of all moulds. A set of finalised mould designs are shown in Fig.~\ref{fig:final_mould_desing_examples}.

\subsection{Contractile network dipole orientation model}

To simulate tissue growth within the moulds, we use a contractile network dipole orientation (CONDOR) model that predicts the self-organisation of cells in polarised tissues \cite{hague2019a}. The form of the model is,
\begin{equation} 
E = \sum_{ij} \frac{\kappa_{0}\bar{\kappa}_{ij}}{2}\left( |\lvec_{ij}| - l_{0,ij}\left(1-\frac{\Delta}{2}\left(2-|\hat{\lvec}_{ij}\cdot\svec_{i}|^2 -|\hat{\lvec}_{ij}\cdot\svec_{j}|^2\right) \right)\right)^2
\end{equation}
The model describes the interaction between elongated (or polarised) cells and the ECM, which is mediated via integrins. $E$ represents the total energy of the system of cells and force dipoles.

The system of cells and ECM is modelled as a contractile network of springs \cite{boal} within which the equilibrium length of the springs is modified to represent cell-matrix interactions. The interactions draw in the ECM perpendicular to the orientation of the polarised cells creating a force dipole. The alignment of cells is represented by the unit vector $\svec$, and the cell induces a force dipole that pulls on the ECM perpendicular to $\svec$. Indices $i$ and $j$ are used to label cells. $\Delta$ is a dimensionless parameter of value between 0 and 1 representing the strength of the cell-matrix interaction. $l_{0}$ is the equilibrium distance of the contractile network segments between nearest-neighbour cells in the absence of cell-matrix interactions. $\lvec_{ij}$ is the displacement between cells $i$ and $j$. $\svec_{i}$ is a unitary vector representing the orientation of cell $i$. The spring constant $\kappa_{0}$ represents the value of the nearest-neighbour spring constant. The dimensionless spring constant for the bond between cells $i$ and $j$ is represented by $\bar{\kappa}_{ij}=\kappa_{ij}/\kappa_{0}$ such that $\bar{\kappa}_{\rm nn}=1$. For bonds between more distant cells $\bar{\kappa}_{ij}<1$. 

The lowest energy state of this model is found using a simulated annealing approach as detailed in Ref. \cite{hague2019a}. During the anneal, cells maintain their bonds to the ECM (and thus their relative positions in the final configuration), but are otherwise free to move. This is consistent with the growth of cellular hydrogels, where the timescales associated with self-organisation are less than those related to cell motility (i.e. on the timescale on which we are interested, cells are not motile \cite{hague2019a}). After the anneal we can calculate additional values, such as the average tension on a cell, $\tau$ or the average orientation, e.g. $s_{x}^2$

The parameters of the simulations are as follows: We run simulations with cells and their bonds initially placed on a face-centre-cubic (FCC) lattice. The FCC lattice ensures that there is non-zero shear modulus in the model (the shear modulus is absent in contractile networks arranged as simple cubic lattices \cite{boal}). In addition, in the case of open boundary conditions, it is possible for the whole lattice to fold about a plane, which can occur during the simulated annealing process. This can be resolved by including more distant bonds in the contractile network representing the ECM, so that folds along linear axes are not possible without significant compression in the contractile network leading to a high energy state. In the simulations relating to this model, we set the dimensionless constants $\Delta=0.3$ for all cells, $\bar{\kappa}_{\rm NNN} = 0.4$, $\bar{\kappa}_{\rm NNNN} = 0.2$ and $\bar{\kappa}_{\rm NNNNN} = 0.2$ (where the subscripts NNN, NNNN and NNNNN refer to next-nearest, next-next-nearest and next-next-next-nearest neighbours respectively). The anneal temperature schedule is exponential with an initial temperature of $T_{\rm init}/\kappa_{0}l_{0}^2=0.0625$. In practical terms, we set $l_{0}=40$ and $\kappa_{0}=1$ such that $T_{\rm init}=100$. A fixed number of iterations is used ($N_{\rm cells} \times 10^{6}$) and the ratio between initial and final temperatures $T_{\rm init}/T_{\rm final} = 10^{7}$. The probability of selecting a new position follows a step-function probability. For the mould sizes selected, $N_{\rm cells}\approx 10000$. A large penalty of $10^{9}$ is applied to the energy for each cell that is situated outside the mould area, essentially forbidding cells to move into such regions.

 We concentrate on moulds here. We expect the method to be valid for other situations such as scaffolds, however moulds offer a convenient system where experimental data tracking self-organisation in cellular hydrogels are available due to the ease of imaging flat systems (albeit data are sparse) \cite{georgiou2013, mukhey2018}.

Mould shapes generated as described in Section \ref{sec:mouldgeneration} are scaled to a suitable size for the CONDOR simulation and encoded in a voxel space. For our training set, the simulation area has an equal height and width of $62.5 l_{0}$; the mould is centred in this area and scaled so that the maximum height or width is equal to $56.24 l_{0}$, or $0.9$ times the simulation area dimensions. The mould is further scaled so its area is exactly 40\% of the total simulation area. When encoded into voxel space format, the mould is given a depth of $4 l_{0}$.

The CONDOR simulation of each individual tethered mould in this set takes approximately 24-36 core hours. We run the simulations in parallel on a multi-core machine. The output of the CONDOR simulation describes the 3D cell lattice in its final state, with specific positions, orientations etc. being listed for each individually simulated cell. The machine learning model used to generate results in this paper was trained using a total of 6500 examples outputs of the CONDOR simulations representing approximately 27 core years of simulation, with 9:1 training and validation split (trained on 5760 examples, validation on 640). We have additionally run the CONDOR simulation for a set of 100 moulds to be set aside as a test set to assess the results of training, along with selection of several mould designs also used for real cell growth.

\subsection{Machine learning}

We use the TensorFlow \cite{tensorflow2015-whitepaper} framework for machine learning to implement the \texttt{pix2pix} conditional generative adversarial network (cGAN) described in \cite{2016arXiv161107004I}. Full details of the model architecture and an explanation of its function can be found in the Appendix. For convenience, we shall refer to our trained model as CONDOR-ML in the following. 



In order to be used in the adversarial model, data output from the CONDOR simulations was converted into a set of 2D fields (flattening the z axis) encapsulating the properties of the cell matrix, which in turn can be encoded within a 256 by 256 pixel image compatible with the machine learning algorithm. The length and width of each pixel scales with equilibrium distance. For convenience, we define this as $l_{p}=125 l_{0} / 512$. The equivalent 3D volume that each pixel represents is $4 l_{p}^2 l_{0}$. 

Cells in the CONDOR simulations are found at specific points, whereas a continuous field is needed for the adversarial model. The average spacing between cells in the $xy$ plane are nominally larger than the sizes of pixels we are using for our distributions. As such a simple binning procedure would mean that many pixels would not contain any cells and consequently no measure of any cell properties. 

To compensate for this and ensure we obtain a smooth and continuous field, we use a method of kernel density estimation to produce a weighted sum of each property ($Z_{p}$) per pixel based on all cells in the surrounding area with
\begin{equation}
\label{eqn:kde_pixel_sum}
Z_{p} = \sum_{i} z_{i} w_{i,p}
\end{equation}
where $z_{i}$ is the value an individual cell has for one of its properties and $w_{i,p}$ is the associated weight. The summation is made over all individual cells. Weights are based on the distance between a cell and pixel and determined using a multivariate normal probability density function calculated over the area of the pixel,
\begin{equation}
\label{eqn:kde_normal_distribution}
w_{i,p} = \int_{x_{p}-l_{p}/2}^{x_{p}+l_{p}/2} \int_{y_{p}-l_{p}/2}^{y_{p}+l_{p}/2} \frac{1}{2 \pi \sigma^{2}} e^{\left[ -\frac{1}{2\sigma^{2}} \left( \left( x-x_{i} \right)^{2} + \left( y-y_{i} \right)^{2} \right) \right]} \drm y \drm x
\end{equation}
where $x_{i}$ and $y_{i}$ are the 2D projections of the coordinates of the cell, $x_{p}$ and $y_{p}$ are the centre coordinates of the pixel; these are all measured or converted to units of pixel length ($l_{p}$). $\sigma$ defines the width of this normal distribution and is also measured on a pixel scale; to ensure the distribution is smooth without losing any detail, i.e. small scale variations in density or other properties, we set $\sigma = 3$px. 

The specific properties of cells from the CONDOR simulations included in the training are average magnitude of tension on a cell ($\tau_{i}$), squares of the cell orientation ($s_{x,i}^{2}$, $s_{y,i}^{2}$, $s_{z,i}^{2}$) and the tensor product of the cell orientations ($s_{x,i}s_{y,i}$, $s_{x,i}s_{z,i}$, $s_{y,i}s_{z,i}$). All measures are dimensionless except $\tau_{i}$, which we convert to unitless values using
\begin{equation}
\tau_{i}=\frac{\tau_{0,i}}{\kappa_0 l_{0}}
\end{equation}
where $\tau_{0,i}$ is the initial total tension on each cell. For the purposes of calculating the cell density field, we also associate a property $\rho_{i}=1$ with each cell (i.e. to calculate the convolution of the cell positions). Both equations~\ref{eqn:kde_pixel_sum} and \ref{eqn:kde_normal_distribution} apply to all cell properties; the corresponding symbols we use for both the individual cell and weighted sum of quantities is shown in Table~\ref{tab:symbol_conversion}. 


It is important to note that as a consequence of this convolution, each field is inherently multiplied by the density distribution for cells and is hence not individually normalised. Property distributions can subsequently be normalised using the inverse of the  mean of the corresponding density distribution, such that
\begin{equation}
\label{eqn:norm_factor_density}
\overline{Z_{p}} = \frac{N_{p}}{\sum_{p} \left( \sum_{i} \rho_{i} w_{i,p} \right)} \sum_{i} z_{i} w_{i,p}
\end{equation}
where $N_{p}$ is the total number of pixels in the distribution (which for the $256$ by $256$ pixel distributions we have used would be $N_{p}=65536$). As we also set $\rho_{i} = 1$, this becomes equivalent to 
\begin{equation}
\overline{Z_{p}} = \frac{N_{p}}{N_{c}} \sum_{i} z_{i} w_{i,p}
\label{eqn:normalisation}
\end{equation}
where $N_{c}$ is the total number of cells, which itself is equal to the sum of weights for all cells and pixels, equivalent also to the sum of values in the density distribution. This normalisation also makes $\overline{Z_{p}}$ a dimensionless quantity for all properties.

\begin{table}[h]
    \centering
    
    \begin{tabular}{c||c|c|c|c|c|c|c|c}
        Quantity & Density & Tension & \multicolumn{6}{c}{Tensor product of orientation} \\
        \hline
        Cell property ($z_{i}$) & $\rho$ & $\tau$ & $s_{x}^{2}$ & $s_{y}^{2}$ & $s_{z}^{2}$ & $s_{x}s_{y}$ & $s_{x}s_{z}$ & $s_{y}s_{z}$ \\
        \hline
        Field ($Z_{p}$) & $P$ & $T$ & $S_{x}^{2}$ & $S_{y}^{2}$ & $S_{z}^{2}$ & $S_{x}S_{y}$ & $S_{x}S_{z}$ & $S_{y}S_{z}$
    \end{tabular}
    \caption{Symbols we use for values taken from CONDOR simulations (cell properties) and the symbols for their corresponding weighted sums (fields).}
    \label{tab:symbol_conversion}
\end{table}

While normalisation of cell property distribution is important for making use of results, it is not applied during training of the machine learning model. Since the total number of cells can vary (and does for individual examples in our training data), this means that the normalisation factor $N_{p}/N_{c}$ is not universal across the training set and ultimately results in a more complex relation that needs to be learned. The machine learning algorithm instead predicts the weighted sums ($Z_{p}$) which can then subsequently be normalised using the inverse of the mean of the corresponding predicted density as per Equation~\ref{eqn:norm_factor_density}.

As a technical note, within the actual training process all distributions for each property in the entire training data set are uniformly re-scaled as to have a value range from $-1$ to $1$. Subsequent predictions made by the trained model produces distributions in this $-1$ to $1$ range, for which we apply the reverse of this scaling to return it to a scale matching the training data.

\begin{figure}[h]
\centering
{
\includegraphics[width=\textwidth]{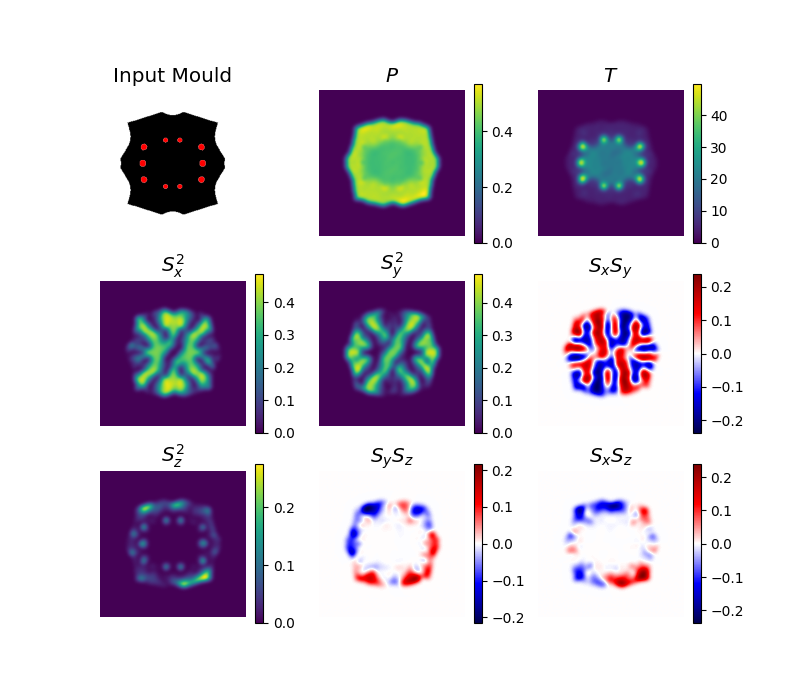}
\caption{An example of the training data used, including the input mould image and all 8 target fields produced by the CONDOR simulation. All properties are dimensionless.}
\label{fig:mould_123_example_training}
}
\end{figure}

Fig.~\ref{fig:mould_123_example_training} shows the final weighted sum distributions for all 8 cell property fields for an example simulation result for a mould and tether arrangement. There are a number of aspects of these distributions common with all results. Firstly, the distributions for squares of cell orientations ($S_{x}^{2}$, $S_{y}^{2}$ and $S_{z}^{2}$) do not sum to a uniform value of $1$; as a result of the kernel density estimation method, the sum of these distributions is instead equal to the cell density distribution $P$. As $S_{z}^{2}$ is typically small in moulds, the $S_{x}^{2}$ and $S_{y}^{2}$ distributions appear as approximate inverses each other. There are increased $S_{z}^{2}$ and tension ($T$) values around the locations of tethers. $S_{z}^{2}$ also increases close to the edges of the cell area. Because a fixed lattice of cells is used as a convenient way of defining the tether within CONDOR simulations, non-zero density is measured at the locations of tethers. The boundary of the density distribution is not sharply defined due to the kernel density estimation. 

The corresponding input image needed for the machine learning algorithm is the associated mould and tethers used in the simulation. This consists of two layers, one describing mould area and the other tether positions. Both layers consist of 256 by 256 arrays of binary values - in the case of the mould layer, values of 1 indicate the depression in the mould where the cell matrix lies, while 0 indicates any region outside this. Similarly, values of 1 in the tether layer indicate the presence of a tether and values are 0 otherwise. Any pixel that is set as a tether location is also set as being non-mould area. The final input image is a 256 by 256 by 2 array.

\section{Results}

Our test data set contains a range of examples covering all types of mould symmetry designs and tether distributions. Here we present results for a selection of different moulds and the comparisons between the results of CONDOR with the machine learning predictions of CONDOR-ML. For the purposes of directly comparing predictions from the machine learning model with simulated counterparts, we do not initially apply the normalisation from Eqn.~\ref{eqn:normalisation}.

\begin{figure}[hp]
\centering
{
\includegraphics[width=0.95\textwidth]{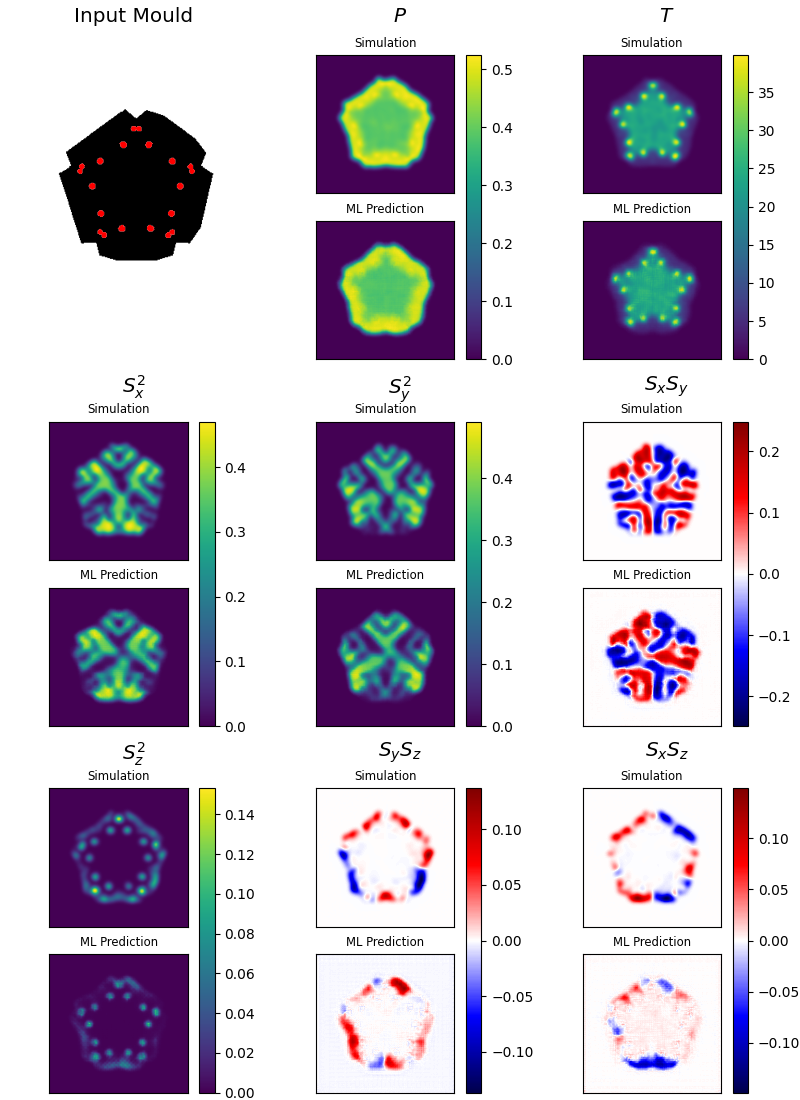}
\caption{Comparison between the results of the CONDOR simulation and the machine learning prediction for a an example mould with 5 order rotational symmetry from our testing data set. All properties are dimensionless.}
\label{fig:mould_12_results_comparisons}
}
\end{figure}

Fig.~\ref{fig:mould_12_results_comparisons} shows comparisons between CONDOR simulations and CONDOR-ML predictions of all properties for an example mould and tether set with 5th order dihedral rotational symmetry from our testing data. The final cell matrix shape is given by the $P$ field, which shows a close match between the CONDOR-ML prediction and the simulation, with only minor deviations in the outline. The overall density distribution itself matches the simulation closely, with CONDOR ML correctly predicting a lower and uniform density in the central area between tethers. The same can be said with tension, where CONDOR-ML has predicted the increased value around tether positions and the average value between them, though fluctuations in the central region are not identical.

Complex $S_{x}^{2}$, $S_{y}^{2}$ and $S_{x} S_{y}$ distributions can be seen in the simulations, and Fig.~\ref{fig:mould_12_results_comparisons} shows that CONDOR-ML predicts these well, despite their complexity. The machine learning prediction correctly identifies that $S_{x}^{2}$ and $S_{y}^{2}$ follow approximately inverse distributions of each other. The most significant mismatches between the CONDOR simulation and machine learning predictions are found in the distributions of $S_{z}^{2}$, $S_{y} S_{z}$ and $S_{x} S_{z}$ (Fig.~\ref{fig:mould_12_results_comparisons}). While increased $S_{z}^{2}$ is correctly predicted around tethers, increases in $S_{z}^{2}$ at the edges of the cell matrix are less consistent. The $S_{y} S_{z}$ and $S_{x} S_{z}$ fields match less well, although the regions where these properties are non-zero are small and typically at the edge of the tissue; we do note that regions where non-zero values occur are predicted correctly even if the values are incorrect.

\begin{figure}[hp]
\centering
\includegraphics[width=\textwidth]{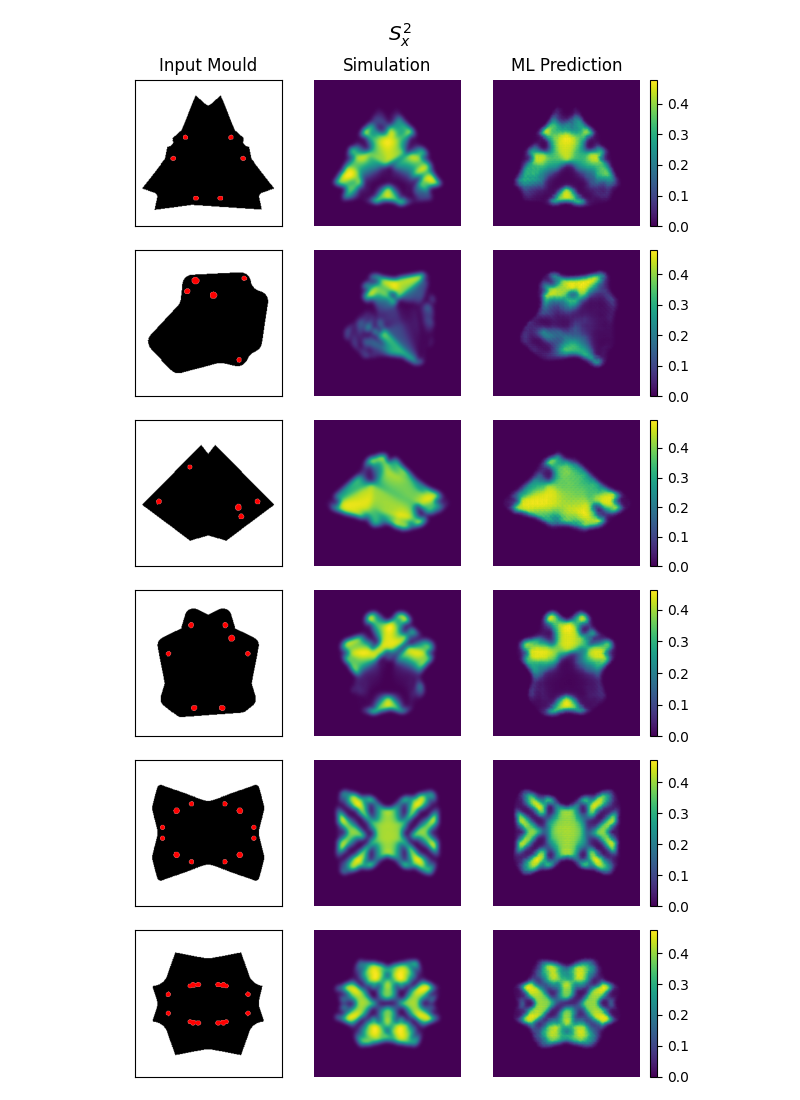}
\caption{Comparisons between the CONDOR simulation results and CONDOR-ML prediction results for $S_{x}^{2}$ with best agreement. All properties are dimensionless.}
\label{fig:sx2_best_agreement_examples_1}
\end{figure}

\begin{figure}[hp]
\centering
\includegraphics[width=\textwidth]{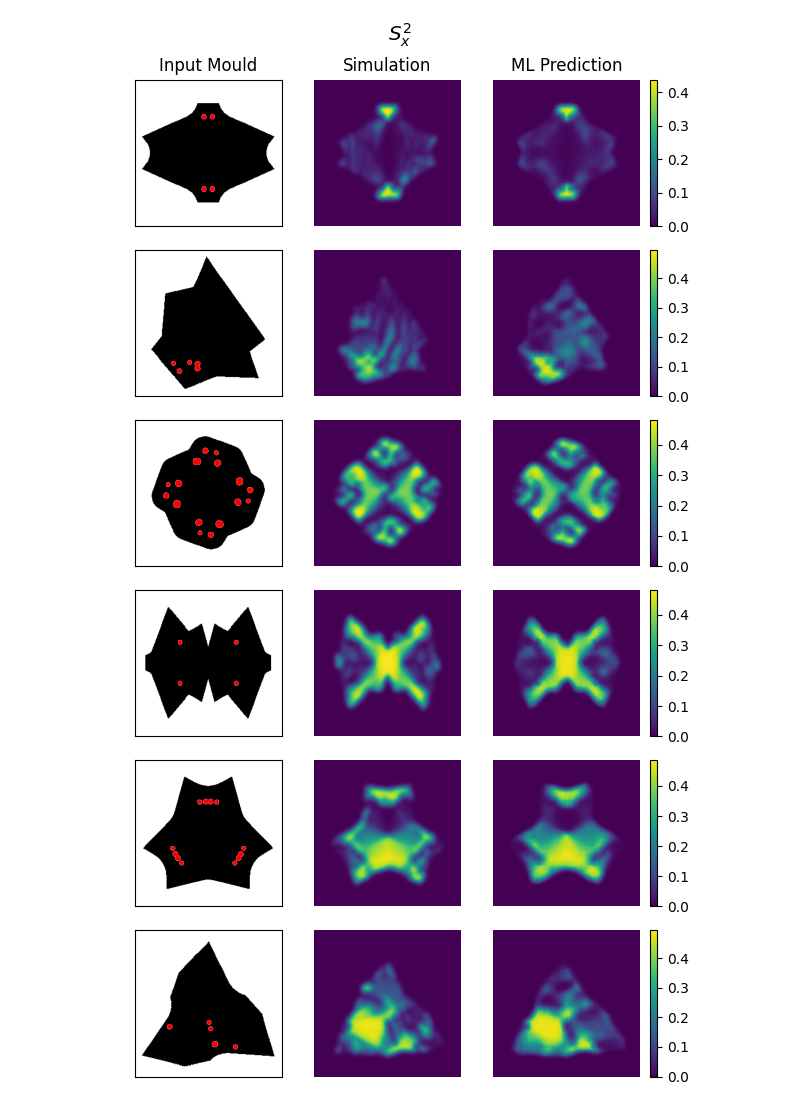}
\caption{Comparisons between the CONDOR simulation results and CONDOR-ML prediction results for $S_{x}^{2}$ with worst agreement. All properties are dimensionless.}
\label{fig:sx2_worst_agreement_examples_1}
\end{figure}

\begin{figure}[hp]
\centering
\includegraphics[width=\textwidth]{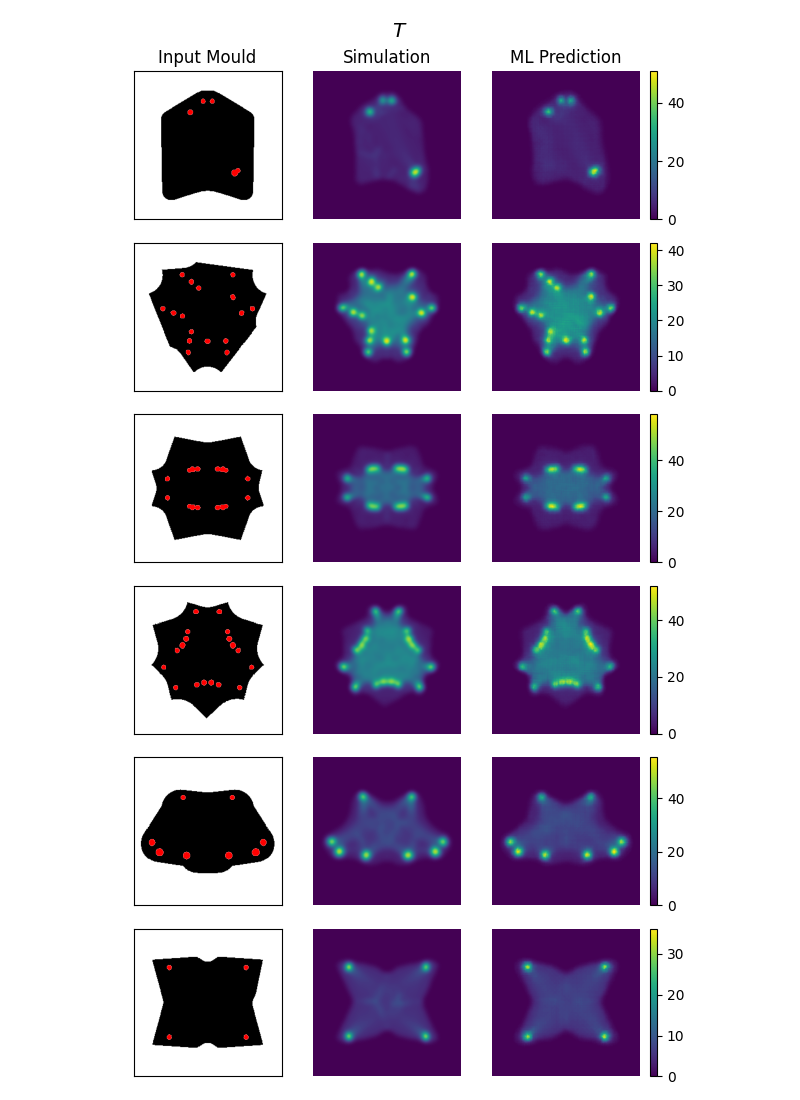}
\caption{Comparisons between the CONDOR simulation results and CONDOR-ML prediction results for $T$ with best agreement. All properties are dimensionless.}
\label{fig:tension_best_agreement_examples_1}
\end{figure}

\begin{figure}[hp]
\centering
\includegraphics[width=\textwidth]{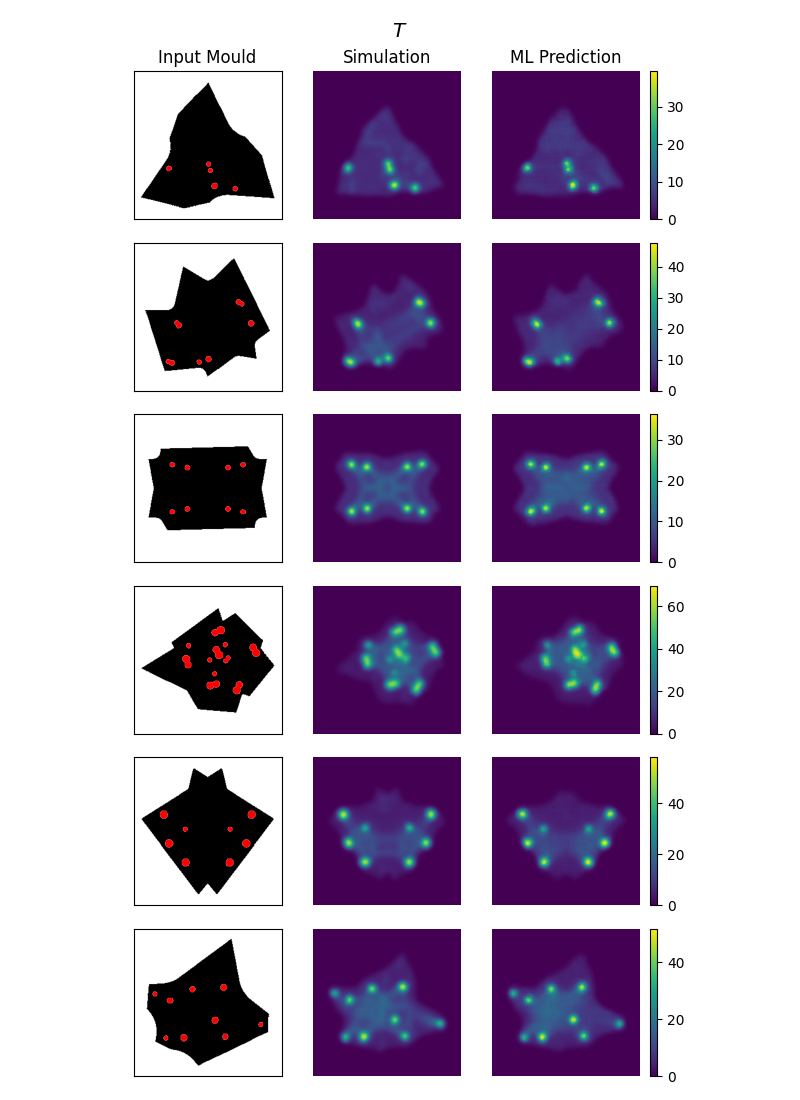}
\caption{Comparisons between the CONDOR simulation results and CONDOR-ML prediction results for $T$ with worst agreement. All properties are dimensionless.}
\label{fig:tension_worst_agreement_examples_1}
\end{figure}

We measure agreement between CONDOR simulations and CONDOR-ML predictions for each property in each mould of our test set using the mean absolute error (MAE),
\begin{equation}
\mathrm{MAE} = \frac{1}{N_{p}} \sum_{p} \left| Z_{p,\mathrm{C}} - Z_{p,\mathrm{ML}} \right|
\end{equation}
where $Z_{p,\mathrm{C}}$ are the pixel values from the simulation and $Z_{p,\mathrm{ML}}$  the corresponding values from the machine-learning prediction. We used this measurement to identify the examples in our test set with the best and worst agreement for the properties $S_{x}^{2}$ and $T$, as these are the most important properties to consider from a rational design standpoint. The 6 best and worst agreements for $S_{x}^{2}$ are shown in Fig.~\ref{fig:sx2_best_agreement_examples_1} and Fig.~\ref{fig:sx2_worst_agreement_examples_1} respectively, while the 6 best and worst for $T$ are shown in Fig.~\ref{fig:tension_best_agreement_examples_1} and Fig.~\ref{fig:tension_worst_agreement_examples_1}. We note that the visual differences between the simulated and predicted distributions of $S_{x}^{2}$ and $T$ are minor even in the worst cases.

\begin{figure}[hp]
\centering
\includegraphics[width=\textwidth]{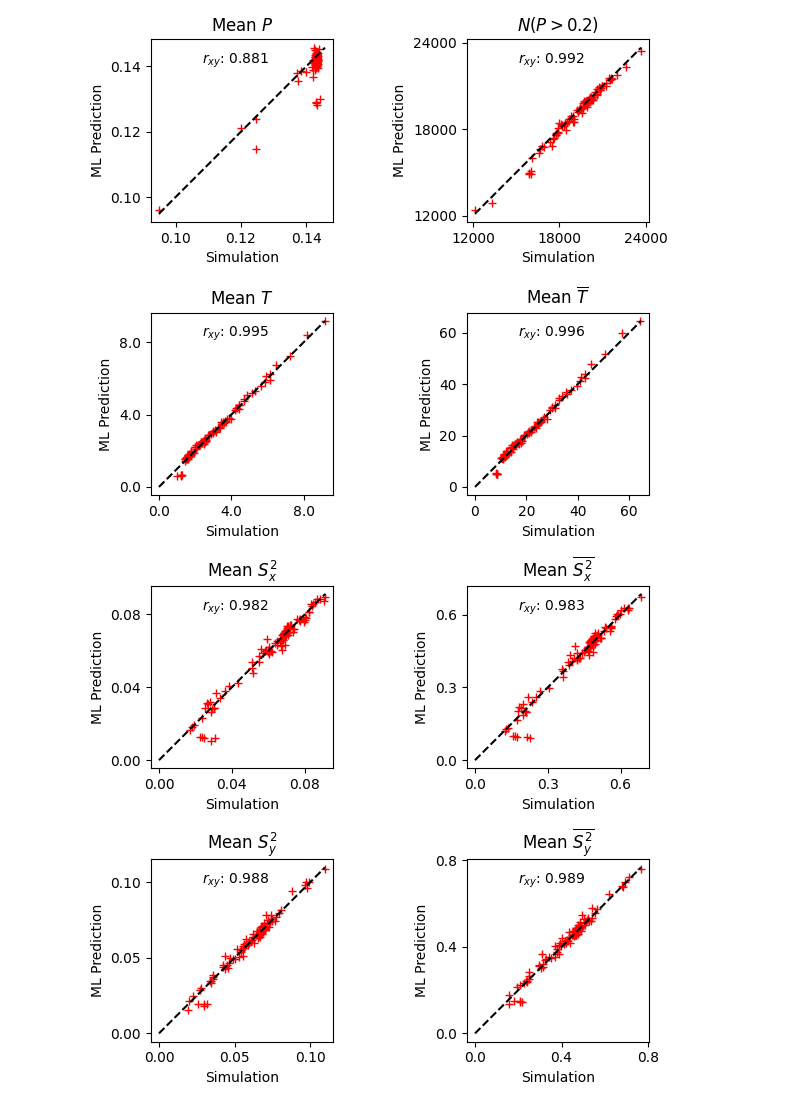}
\caption{Figure showing statistical comparisons for the entire test set. Statistics include mean pixel density $P$, number of pixels with density exceeding $0.2$ as a proxy for area, $N \left( P > 0.2 \right)$, and mean values for both initial and normalized tension ($T$ and $\overline{T}$), $x$ alignment ($S_{x}^{2}$ and $\overline{S_{x}^{2}}$) and $y$ alignment ($S_{y}^{2}$ and $\overline{S_{y}^{2}}$). Also shown for each comparison is the Pearson correlation coefficient, $r_{xy}$, and a dashed line to indicate equal values. All properties are dimensionless. Barred and unbarred properties are related by a scaling factor depending on $N_{c}$. Note that the the axes of the top two plots do not start at 0.}
\label{fig:statistics_comparisons_1}
\end{figure}

We also measure the accuracy of predictions against expected simulated results for the entire test set. This is done by calculating a statistical measure of a particular property (for example, mean density) for both the predicted and simulated result of each example and plotting these values against each other, as shown in Figure~\ref{fig:statistics_comparisons_1}. For each comparison we also calculate the Pearson correlation coefficient ($r_{xy}$) using
\begin{equation}
r_{xy}=\frac{ N \sum_{i} X_{i} Y_{i} -  \sum_{i} X_{i} \sum_{i} Y_{i} }{ \sqrt{N \sum_{i} X_{i}^{2} - \left( 
\sum_{i} X_{i} \right)^{2}} \sqrt{N \sum_{i} Y_{i}^{2} - \left( 
\sum_{i} Y_{i} \right)^{2}} }
\end{equation}
where $X_{i}$ represents statistics from simulated results, and $Y_{i}$ the corresponding statistics from the machine learning predicted results, and $N$ is the total number of examples in our test data set.

Comparisons made in Figure~\ref{fig:statistics_comparisons_1} include mean density, which as noted previously is used as the scale factor for normalisation. This shows a mostly good correlation, confirmed with $r_{xy}=0.881$, though with 5 outliers which have a predicted mean density that is approximately 7\% lower than the corresponding values from simulations; these outliers arise from localised variations in density predictions for tapered and intricate portions of mould shapes where no tethers are present. Such cases are unlikely to be used for most tissue engineering applications.

Additional statistical comparisons made in Figure~\ref{fig:statistics_comparisons_1} include the number of pixels with density exceeding $0.2$, which effectively is a measure of area of the cell matrix. We also compare the means for both initial and density-normalised tension and cell orientations on the $x$ and $y$ axes. All statistics show high agreement, with $r_{xy}$ values all exceeding $0.98$. The outliers associated with cell orientation are the same as those associated with the mean density.

The normalised properties $\overline{T}$, $\overline{S_{x}^{2}}$ and $\overline{S_{y}^{2}}$ also show good agreement. When calculating average quantities directly from CONDOR simulation results it is the barred quantities that are calculated. Hence this shows that real quantities, such as total average alignment, which may be useful for design purposes, can be predicted using the machine learning algorithm.


\section{Conclusions and Outlook}

In this paper we have shown proof-of-concept for the use of machine learning tools to predict the self-organisation of cell-laden hydrogels grown in moulds. The machine learning method was trained using the results of contractile-network-dipole orientation models, so makes predictions of self-organisation driven by cell-matrix interactions. 

To create a large training set for the machine-learning algorithm, we developed an automated process for the generation of mould designs, and used a biophysical model to simulate the cell-matrix interactions within those moulds. In this way, we created a large training set with 1000s of cases describing the self-organisation of cell-laden hydrogels within these moulds. These data are used to train an implementation of the \texttt{pix2pix} deep learning model.

The machine learning algorithm has high accuracy and is significantly faster than the biophysical method, opening the possibility of very-high-throughput rational designs for moulds for pharmaceutical testing, regenerative medicine and fundamental studies of biology. Accuracy was tested using $100$ simulations that were unseen in the training of the neural network, demonstrating  that the machine learning algorithm made excellent predictions that agreed well with the CONDOR simulations. Our \texttt{pix2pix} model runs at least 10,000 times faster than a typical CONDOR run. It executes in $\sim0.1$ seconds on a typical laptop CPU or $\sim0.01$ seconds if a laptop-grade GPU is used for the computation. 

We anticipate a design process in which the CONDOR-ML machine learning algorithm is used to make rapid predictions of the tissue self-organisation in tethered mould designs. Once candidate designs were selected in this way, they would be tested by using full CONDOR simulations to confirm the self-organisation within the tissue matched the results from CONDOR-ML. Since the biophysical simulations are also approximate, the process would be finalised by 3D printing the best moulds and growing cultured tissues within the mould.


The advantage of CONDOR-ML over biophysical models and direct growth of biological samples is speed. Disadvantages are that results are approximate (biological growth is definitive) and that the machine learning algorithm is currently only trained on tethered moulds (extension of the machine learning algorithm to 3D requires an entirely new training set, whereas CONDOR is immediately applicable to 3D). Future extensions for scaffolds and 3D bioprinting will open additional applications. The results described here are primarily validated against hydrogels of glial cells, but we expect applicability to any elongated cell types that self-organise by aligning due to the symmetries underlying the model. For fully three dimensional tissue, the speed of full CONDOR calculations limits the size of the system that can be simulated. In full 3D, we are able to simulate systems with resolution of approximately 46 cells along each axis, which limits the feature size available in simulations for e.g. scaffolds (examples of 3D scaffolds can be found in Ref. \cite{loh2013a}). Extensions of this machine learning technique could allow predictions to be made for larger scaffolds and work in this direction is currently in progress. Finally, we believe that the machine learning techniques used here could be applied to other biophysical models of self-organisation in tissue, or directly to experimental data, if sufficient data could be gathered.

\section*{Acknowledgments}

We wish to thank James Phillips, Richard Blythe and Elsen Thjung for useful discussions. We acknowledge funding from the STFC Impact Acceleration Account.



\section*{References}

\bibliographystyle{iopart-num}
\bibliography{references}

\providecommand{\newblock}{}
\begin{thebibliography}{10}
\expandafter\ifx\csname url\endcsname\relax
  \def\url#1{{\tt #1}}\fi
\expandafter\ifx\csname urlprefix\endcsname\relax\def\urlprefix{URL }\fi
\providecommand{\eprint}[2][]{\url{#2}}

\bibitem{kular2014a}
Kular J~K, Basu S and Sharma R~I 2014 {\em Journal of Tissue Engineering\/}
  {\bf 5} 2041731414557112

\bibitem{Hayrapetyan2020}
Hayrapetyan L and Sarvazyan N 2020 {\em Extracellular Matrix and Adhesion
  Molecules\/} (Cham: Springer International Publishing) pp 29--38

\bibitem{bajaj2014a}
Bajaj P, Schweller R~M, Khademhosseini A, West J~L and Bashir R 2014 {\em
  Annual Review of Biomedical Engineering\/} {\bf 16} 247--276

\bibitem{weinhart2019a}
Weinhart M, Hocke A, Hippenstiel S, Kurreck J and Hedtrich S 2019 {\em
  Pharmacological Research\/} {\bf 139} 446--451

\bibitem{jensen2018a}
Jensen G, Morrill C and Huang Y 2018 {\em Acta Pharmaceutica Sinica B\/} {\bf
  8} 756--766

\bibitem{benarye2019a}
Ben-Arye T and Levenberg S 2019 {\em Frontiers in Sustainable Food Systems\/}
  {\bf 3}

\bibitem{capel2019a}
Capel A~J, Rimington R~P, Fleming J~W, Player D~J, Baker L~A, Turner M~C, Jones
  J~M, Martin N~R~W, Ferguson R~A, Mudera V~C and Lewis M~P 2019 {\em Frontiers
  in Bioengineering and Biotechnology\/} {\bf 7}

\bibitem{georgiou2013}
Georgiou M, Bunting S~C, Davies H~A, Loughlin A~J, Golding J~P and Phillips J~B
  2013 {\em Biomaterials\/} {\bf 34} 7335--43

\bibitem{mukhey2018}
Mukhey D, Phillips J, Daniels J and Kureshi A 2018 {\em Acta Biomaterialia\/}
  {\bf 67} 229

\bibitem{hague2019a}
Hague J~P, Mieczkowski P~W, O’Rourke C, Loughlin A~J and Phillips J~B 2020
  {\em Phys. Rev. Res.\/} {\bf 2} 043217

\bibitem{silvanus2017}
Silvanus A, Poulami G and Guillaume S 2017 {\em Phil. Trans. R. Soc. B\/} {\bf
  372} 20150520

\bibitem{bi2016}
Bi D, Yang X, Marchetti M~C and Manning M~L 2016 {\em Phys. Rev. X\/} {\bf
  6}(2) 021011

\bibitem{deshpande2006}
Deshpande V, McMeeking R and Evans A 2006 {\em Proc Natl Acad Sci U S A\/} {\bf
  103} 14015–14020

\bibitem{pathak2008}
Pathak A, Deshpande V, McMeeking R and Evans A 2008 {\em J. R. Soc.
  Interface\/} {\bf 5} 507--524

\bibitem{obbinkhuizer2014}
Obbink-Huizer C, Foolen J, Oomens C, Borochin M, Chen C, Bouten C and Baaijens
  F 2014 {\em Biomech. Model. Mechanobiol.\/} {\bf 13} 1053--1063

\bibitem{legand2009}
Legant W, Pathak A, Yang M, Deshpande V, McMeeking R and Chen C 2009 {\em Proc.
  Natl. Acad. Sci. USA\/} {\bf 106} 10097–10102

\bibitem{schwarz2013}
Schwarz U~S and Safran S~A 2013 {\em Rev. Mod. Phys.\/} {\bf 85}(3) 1327--1381

\bibitem{2021MNRAS.504.2603T}
{Thorne} B, {Knox} L and {Prabhu} K 2021 {\em \mnras\/} {\bf 504} 2603--2613

\bibitem{2022arXiv220613306D}
{Demianenko} M, {Samorodova} E, {Sysak} M, {Shiriaev} A, {Malanchev} K,
  {Derkach} D and {Hushchyn} M 2023 Supernova light curves approximation based
  on neural network models {\em Journal of Physics: Conference Series\/} vol
  2438 (IOP Publishing) p 012128

\bibitem{frid2018gan}
Frid-Adar M, Diamant I, Klang E, Amitai M, Goldberger J and Greenspan H 2018
  {\em Neurocomputing\/} {\bf 321} 321--331

\bibitem{wolterink2017deep}
Wolterink J~M, Dinkla A~M, Savenije M~H, Seevinck P~R, van~den Berg C~A and
  I{\v{s}}gum I 2017 Deep mr to ct synthesis using unpaired data {\em
  Simulation and Synthesis in Medical Imaging: Second International Workshop,
  SASHIMI 2017, Held in Conjunction with MICCAI 2017, Qu{\'e}bec City, QC,
  Canada, September 10, 2017, Proceedings 2\/} (Springer) pp 14--23

\bibitem{2020RemS...12..901D}
{de Bem} P~P, {de Carvalho Junior} O~A, {Fontes Guimar{\~a}es} R and {Trancoso
  Gomes} R~A 2020 {\em Remote Sensing\/} {\bf 12} 901

\bibitem{2020TCry...14..565B}
{Bolibar} J, {Rabatel} A, {Gouttevin} I, {Galiez} C, {Condom} T and {Sauquet} E
  2020 {\em The Cryosphere\/} {\bf 14} 565--584

\bibitem{2020NatSR..10.1322B}
{Ban} Y, {Zhang} P, {Nascetti} A, {Bevington} A~R and {Wulder} M~A 2020 {\em
  Scientific Reports\/} {\bf 10} 1322

\bibitem{2018RSEnv.204..509B}
{Belgiu} M and {Csillik} O 2018 {\em Remote Sensing of Environment\/} {\bf 204}
  509--523

\bibitem{2016arXiv161107004I}
{Isola} P, {Zhu} J~Y, {Zhou} T and {Efros} A~A 2017 Image-to-image translation
  with conditional adversarial networks {\em 2017 {IEEE} Conference on Computer
  Vision and Pattern Recognition, {CVPR} 2017, Honolulu, HI, USA, July 21-26,
  2017\/} ({IEEE} Computer Society) pp 5967--5976

\bibitem{ten.tea.2022.0128}
{Guo} J~L, {Januszyk} M and {Longaker} M~T 2023 {\em Tissue Engineering Part
  A\/} {\bf 29} 2--19 pMID: 35943870

\bibitem{boal}
Boal D~H 2012 {\em Mechanics of the Cell\/} 2nd ed (Cambridge University Press,
  Cambridge, UK)

\bibitem{tensorflow2015-whitepaper}
Abadi M, Agarwal A, Barham P, Brevdo E, Chen Z, Citro C, Corrado G~S, Davis A,
  Dean J, Devin M, Ghemawat S, Goodfellow I, Harp A, Irving G, Isard M, Jia Y,
  Jozefowicz R, Kaiser L, Kudlur M, Levenberg J, Man\'{e} D, Monga R, Moore S,
  Murray D, Olah C, Schuster M, Shlens J, Steiner B, Sutskever I, Talwar K,
  Tucker P, Vanhoucke V, Vasudevan V, Vi\'{e}gas F, Vinyals O, Warden P,
  Wattenberg M, Wicke M, Yu Y and Zheng X 2015 {TensorFlow}: Large-scale
  machine learning on heterogeneous systems software available from
  tensorflow.org \urlprefix\url{https://www.tensorflow.org/}

\bibitem{loh2013a}
Loh Q and Choong C 2013 {\em Tissue Eng Part B Rev.\/} {\bf 19} 485--502

\bibitem{2015arXiv150504597R}
{Ronneberger} O, {Fischer} P and {Brox} T 2015 U-net: Convolutional networks
  for biomedical image segmentation {\em Medical Image Computing and
  Computer-Assisted Intervention - {MICCAI} 2015 - 18th International
  Conference Munich, Germany, October 5 - 9, 2015, Proceedings, Part {III}\/}
  ({\em Lecture Notes in Computer Science\/} vol 9351) ed {Navab} N,
  {Hornegger} J, {Wells} W~M~I and F F~A (Springer) pp 234--241

\bibitem{kingma2017adam}
{Kingma} D~P and {Ba} J 2015 Adam: {A} method for stochastic optimization {\em
  3rd International Conference on Learning Representations, {ICLR} 2015, San
  Diego, CA, USA, May 7-9, 2015, Conference Track Proceedings\/} ed {Bengio} Y
  and {LeCun} Y

\end{thebibliography}

\section*{Appendix: Machine learning details}

We use the TensorFlow \cite{tensorflow2015-whitepaper} framework for machine learning to implement the \texttt{pix2pix} conditional generative adversarial network (cGAN) described in \cite{2016arXiv161107004I}. 
Unconditioned generative adversarial networks (GANs) comprise two separate convolutional neural networks (CNNs) which are referred to as the ``generator'' network and the ``discriminator'' network. The training data are a set of images drawn from a particular domain of interest e.g. animals, motor vehicles, specific types of landscape, or human faces. During training, the generator network learns to map random input vectors into realistic two dimensional images that are consistent with the target domain. At the same time, the discriminator network learns to distinguish real images, drawn from the input training data, from ``fake'' images produced by the generator in response to random input vectors. The training objective for the generator network is to produce images that the discriminator will always classify as real, and the training objective for the discriminator is to always distinguish correctly between real and fake images. Simultaneous optimisation of these two objectives results in a generator network that produces images that are almost indistinguishable from real images in the target domain. 

Conceptually, the \texttt{pix2pix} model differs from standard GANs in two key ways: 

\begin{enumerate}
    \item Firstly, the inputs to the generator network are images rather than random vectors. The training data comprise pairs of input and target images that are related to each other by a shared, unknown mapping function. During training, the objective  of the generator network is to learn this mapping from the input image space to the target image space, while simultaneously producing realistic images that the discriminator network cannot distinguish from the target images.
    \item Secondly, rather than considering its input images in their entirety, the discriminator network is designed to evaluate the realism of small $70\times70$ pixel patches of the generated and target images. Empirically, this has been shown to result in finer details in images produced by the fully trained generator network \cite{2016arXiv161107004I}. 
\end{enumerate}

\begin{figure}
    \centering
    \includegraphics[width=0.8\textwidth]{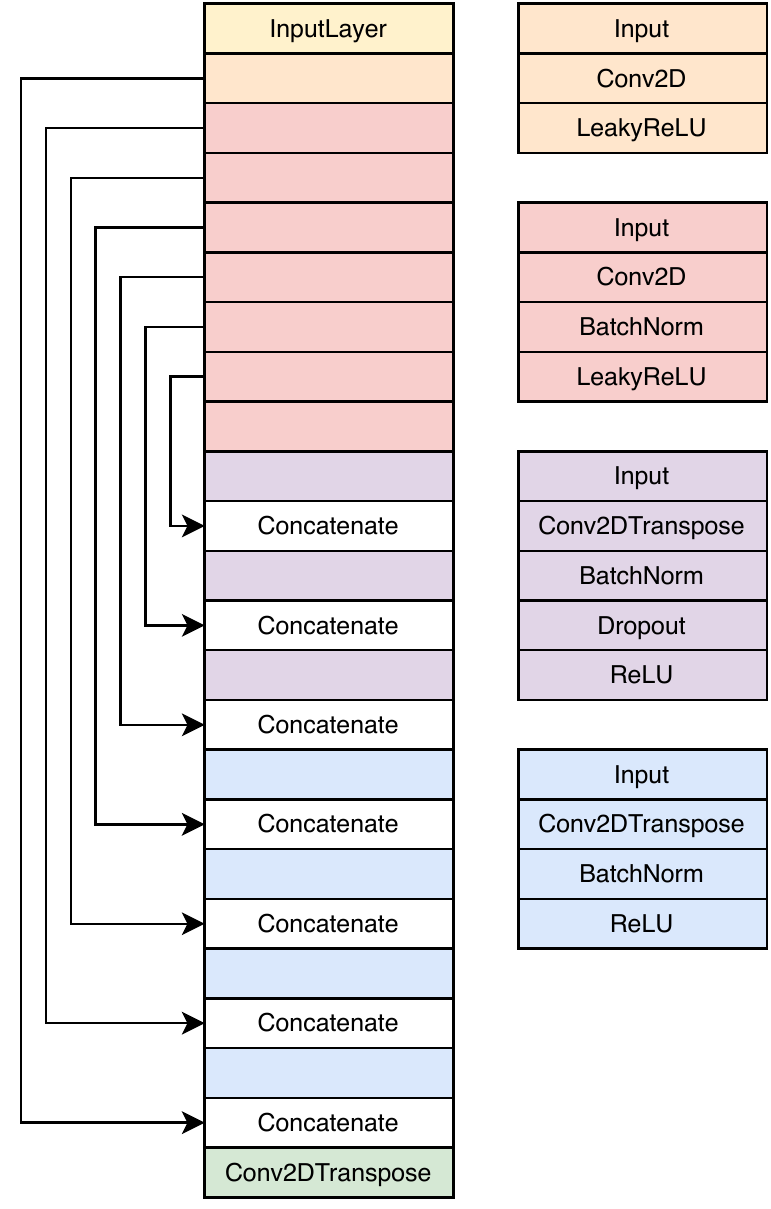}
    \caption{\textit{Left:} Schematic illustration of the \texttt{pix2pix} generator network. Similarly coloured blocks represent blocks containing layers that perform similar, but not necessarily identical, computational operations on their inputs. \textit{Right:} More detailed representations showing the layers that comprise each of the correspondingly coloured blocks in the network schematic. See Table \ref{tab:generator_model} for details of any differences between layers in different computational blocks. ``Concatenate'' blocks stack their inputs along the depth (channel) axis and output the result. Accordingly both inputs to a concatenate layer must have the same width and height, but their channel counts (depths) may be different.}
    \label{fig:generator_model}
\end{figure}

The generator of the \texttt{pix2pix} network uses the U-net \cite{2015arXiv150504597R} architecture shown in Figure \ref{fig:generator_model} with more details in Table \ref{tab:generator_model}. A U-net is a variant of a CNN architecture known as a convolutional auto-encoder. These networks pass their inputs through a sequence of convolution-normalisation-activation blocks (layers 2-24 in Table \ref{tab:generator_model}) called the ``encoder'' network. The encoder extracts features from the input images and encodes them as a one-dimensional array of values called the ``latent space vector'' or ``bottleneck'' (the output of layer 24 in Table \ref{tab:generator_model}). The values in the latent space vector are then passed into a sequence of transpose-convolution-normalisation-activation blocks (layers 25-56 in Table \ref{tab:generator_model}) called the ``decoder'' network. The decoder uses the values in the latent space vector to generate an output image that has the same width and height as the encoder input, but may have a different depth (typically refered to as the number of channels). Three of the blocks in the decoder network also include dropout layers (layers 27, 32 and 37 in Table \ref{tab:generator_model}) which randomly disable a fraction of the weights in the block and introduce noise that helps the entire network to generalise well when supplied with inputs that are not in the training set. During training, the layers of an auto-encoder learn weights such that the bottleneck contains a compressed representation of the input image that contains all the information that the decoder network required to reconstruct the required output image.

The U-net architecture augments a standard auto-encoder by adding ``skip connections'' that pass data directly between layers in the encoder and decoder and bypassing the bottleneck. The authors of \cite{2015arXiv150504597R} showed that the skip connections helped to preserve the overall structure of the input image while allowing fine details to be modified. This behaviour is ideal for our application because the skip connections let the network preserve the shape of the moulds while the weights of the network learn to predict the behaviour of the cells they contain.

The discriminator of the \texttt{pix2pix} network implements the architecture shown in Figure \ref{fig:generator_model} with more details in Table \ref{tab:discriminator_model}. The discriminator is a traditional convolutional neural network, consisting of a sequence of blocks that perform convolution, normalisation, activation and padding operations. The network is designed to ensure that each pixel of its output represents a features of a restricted $70\times70$ pixel region or ``patch'' of the discriminator input. 

As described in Section \ref{sec:mouldgeneration}, we train CONDOR-ML using training examples that comprise a two-channel input image, and a corresponding eight-channel ``label'' image. The input images encode the spatial extent of a randomly generated mould and the locations of any tethers it contains. The labels have the same width and height as the inputs and encode the distributions of cell properties within the mould predicted by the CONDOR code. For each training example, we use the two-channel images as input to the generator and predict an eight-channel output that we will hereafter refer to as the ``generated'' image. 

We then run the discriminator twice, using two different input prescriptions. In the first run, we concatenate the training input image (i.e. an image of the mould) with the corresponding label for that input (i.e. the ``true'' distributions of cell properties within the mould predicted by the CONDOR code). The resultant discriminator output is a single channel $30\times30$ image, that we will call the ``real discriminator response''. In the second run, we concatenate the training input image with the corresponding generated image (i.e. the predicted distributions of cell properties within the mould). The output is another single channel $30\times30$ image, that we will call the ``generated discriminator response''.

We use the three output images to compute an overall ``loss'' that is decreases as the the generated image becomes more consistent with the target domain (i.e. more realistic) and as the mean pixel-wise difference between the generated and label images decreases. The overall loss is composed from several constituent loss functions that each operate on to the outputs from different parts of the \texttt{pix2pix} model. 

Firstly, we compute the mean of the absolute differences between each pixel in the generated and target images. This is conventionally referred to as the ``L1 loss'':
\begin{equation}
    \mathcal{L}_{\mathrm{L1}} = \frac{1}{256^{2}\times8}\sum_{i,j=1}^{256}\sum_{k=1}^{8}|G_{i,j,k} - L_{i,j,k}|
\end{equation}
where $G_{i,j,k}$ and $L_{i,j,k}$ represent the pixels in the $i$th rows, $j$th columns and $k$th channels of the generated and label images, respectively.

As well as the L1 loss we also compute two loss functions based on the discriminator outputs. For the real discriminator response we compute the sigmoid cross-entropy between it and a $30\times30$ array of ones. Similarly we compute the sigmoid cross-entropy between generated discriminator response and a $30\times30$ array of zeros. The sigmoid cross entropy between a prediction $X$ and a target $Y$ is defined as:
\begin{equation}
    \mathcal{L}_{\mathrm{LSXE}}(X, Y) = (Y-1)\log(1 - \mathrm{sigmoid}(X))-Y\log(\mathrm{sigmoid}(X))  
\end{equation}
The sigmoid function is used to map the discriminator network outputs from a potentially infinite domain into the range $[0,1]$.
\begin{equation}
    \mathrm{sigmoid}(x) = \frac{1}{1+e^{-x}}
\end{equation}
To train the discriminator network we then compute the discriminator loss $\mathcal{L}_{\mathrm{disc}}$ as the sum:
\begin{equation}
    \mathcal{L}_{\mathrm{disc}} = \sum_{i, j = 1}^{30}\mathcal{L}_{\mathrm{LSXE}}(D_{i, j}^{L}, 1) + \mathcal{L}_{\mathrm{LSXE}}(D_{i, j}^{G}, 0)
\end{equation}
where $D_{i, j}^{L}$ and $D_{i, j}^{G}$ represent the pixels in the $i$th rows and $j$th columns of the real discriminator response and the generated discriminator response, respectively. As training progresses, the weights of the discriminator network are adjusted by gradient descent to minimise the value of $\mathcal{L}_{\mathrm{disc}}$. This loss is be minimised and equal to zero if the discriminator network predicts an array of zeros for the generated image and an array of ones for the label image.

To train the generator, we compute the generator loss $\mathcal{L}_{\mathrm{gen}}$ as the weighted sum:
\begin{equation}
    \mathcal{L}_{\mathrm{gen}} = \lambda\mathcal{L}_{\mathrm{L1}} + \mathcal{L}_{\mathrm{LSXE}}(D_{i, j}^{G}, 0)
\end{equation}
where $\lambda$ is a model hyper-parameter that we set to 1000, following the authors of \cite{2016arXiv161107004I}. As training progresses, the weights of the generator network are adjusted by gradient descent to minimise the value of $\mathcal{L}_{\mathrm{gen}}$. This loss is minimised when a generated image is identical to the corresponding label image, but it will not be zero unless the discriminator has learned to perfectly distinguish real and generate images. 

Our final model is trained using 5760 training examples, each comprising an input image and a corresponding label image. We train our model for 10 epochs, so the network learns from each example 10 times although the order in which the examples are presented to the network is randomised. When training our model, we update the generator and discriminator weights using gradient descent after computing $\mathcal{L}_{\mathrm{gen}}$ and $\mathcal{L}_{\mathrm{disc}}$ for a single image i.e. we use a batch size of one. For both networks, we use the Adam optimizer \cite{kingma2017adam} with a learning rate of $2\times10^{-4}$ and exponential decay rates $\beta_{1}=0.5$ and $\beta_{2}=0.999$. 

\begin{longtable}{lllr}
    \caption{The architecture of the \texttt{pix2pix} generator network. Each row in the table corresponds with a single functional ``layer'' in the network. The ``Type'' column lists the names used by TensorFlow to define the computational objects comprising the network. The ``Output Shape'' column indicates the shape of the numerical array that results from the computation performed by the array on its inputs. We indicate the zeroth dimension of these shapes as a user-definable batch size, but for the work in this paper this batch size is always one. The ``Parameter Count'' column lists the numbers of parameters (weights) that can be trained for each layer. Horizontal lines divide the table into separate blocks that are illustrated in Figure \ref{fig:generator_model}.}
    \label{tab:generator_model}
\\
\hline
Layer & Type & Output Shape & Parameter Count \\
\hline
1 & InputLayer & (Batch Size, 256, 256, 2) & 0 \\
\hline
2 & Conv2D & (Batch Size, 128, 128, 64) & 2048 \\
3 & LeakyReLU & (Batch Size, 128, 128, 64) & 0 \\
\hline
4 & Conv2D & (Batch Size, 64, 64, 128) & 131072 \\
5 & BatchNormalization & (Batch Size, 64, 64, 128) & 512 \\
6 & LeakyReLU & (Batch Size, 64, 64, 128) & 0 \\
\hline
7 & Conv2D & (Batch Size, 32, 32, 256) & 524288 \\
8 & BatchNormalization & (Batch Size, 32, 32, 256) & 1024 \\
9 & LeakyReLU & (Batch Size, 32, 32, 256) & 0 \\
\hline
10 & Conv2D & (Batch Size, 16, 16, 512) & 2097152 \\
11 & BatchNormalization & (Batch Size, 16, 16, 512) & 2048 \\
12 & LeakyReLU & (Batch Size, 16, 16, 512) & 0 \\
\hline
13 & Conv2D & (Batch Size, 8, 8, 512) & 4194304 \\
14 & BatchNormalization & (Batch Size, 8, 8, 512) & 2048 \\
15 & LeakyReLU & (Batch Size, 8, 8, 512) & 0 \\
\hline
16 & Conv2D & (Batch Size, 4, 4, 512) & 4194304 \\
17 & BatchNormalization & (Batch Size, 4, 4, 512) & 2048 \\
18 & LeakyReLU & (Batch Size, 4, 4, 512) & 0 \\
\hline
19 & Conv2D & (Batch Size, 2, 2, 512) & 4194304 \\
20 & BatchNormalization & (Batch Size, 2, 2, 512) & 2048 \\
21 & LeakyReLU & (Batch Size, 2, 2, 512) & 0 \\
\hline
22 & Conv2D & (Batch Size, 1, 1, 512) & 4194304 \\
23 & BatchNormalization & (Batch Size, 1, 1, 512) & 2048 \\
24 & LeakyReLU & (Batch Size, 1, 1, 512) & 0 \\
\hline
25 & Conv2DTranspose & (Batch Size, 2, 2, 512) & 4194304 \\
26 & BatchNormalization & (Batch Size, 2, 2, 512) & 2048 \\
27 & Dropout & (Batch Size, 2, 2, 512) & 0 \\
28 & ReLU & (Batch Size, 2, 2, 512) & 0 \\
\hline
29 & Concatenate with 21 & (Batch Size, 2, 2, 1024) & 0 \\
\hline
30 & Conv2DTranspose & (Batch Size, 4, 4, 512) & 8388608 \\
31 & BatchNormalization & (Batch Size, 4, 4, 512) & 2048 \\
32 & Dropout & (Batch Size, 4, 4, 512) & 0 \\
33 & ReLU & (Batch Size, 4, 4, 512) & 0 \\
\hline
34 & Concatenate with 18 & (Batch Size, 4, 4, 1024) & 0 \\
\hline
35 & Conv2DTranspose & (Batch Size, 8, 8, 512) & 8388608 \\
36 & BatchNormalization & (Batch Size, 8, 8, 512) & 2048 \\
37 & Dropout & (Batch Size, 8, 8, 512) & 0 \\
38 & ReLU & (Batch Size, 8, 8, 512) & 0 \\
\hline
39 & Concatenate with 15 & (Batch Size, 8, 8, 1024) & 0 \\
\hline
40 & Conv2DTranspose & (Batch Size, 16, 16, 512) & 8388608 \\
41 & BatchNormalization & (Batch Size, 16, 16, 512) & 2048 \\
42 & ReLU & (Batch Size, 16, 16, 512) & 0 \\
\hline
43 & Concatenate with 12 & (Batch Size, 16, 16, 1024) & 0 \\
\hline
44 & Conv2DTranspose & (Batch Size, 32, 32, 256) & 4194304 \\
45 & BatchNormalization & (Batch Size, 32, 32, 256) & 1024 \\
46 & ReLU & (Batch Size, 32, 32, 256) & 0 \\
\hline
47 & Concatenate with 9 & (Batch Size, 32, 32, 512) & 0 \\
\hline
48 & Conv2DTranspose & (Batch Size, 64, 64, 128) & 1048576 \\
49 & BatchNormalization & (Batch Size, 64, 64, 128) & 512 \\
50 & ReLU & (Batch Size, 64, 64, 128) & 0 \\
\hline
51 & Concatenate with 6 & (Batch Size, 64, 64, 256) & 0 \\
\hline
52 & Conv2DTranspose & (Batch Size, 128, 128, 64) & 262144 \\
53 & BatchNormalization & (Batch Size, 128, 128, 64) & 256 \\
54 & ReLU & (Batch Size, 128, 128, 64) & 0 \\
\hline
55 & Concatenate with 3 & (Batch Size, 128, 128, 128) & 0 \\
\hline
56 & Conv2DTranspose & (Batch Size, 256, 256, 8) & 16392 \\
\hline
\end{longtable}

\begin{table}[]
    \centering
    \caption{The architecture of the \texttt{pix2pix} discriminator network. Each row in the table corresponds with a single functional ``layer'' in the network. The ``Type'' column lists the names used by TensorFlow to define the computational objects comprising the network. The ``Output Shape'' column indicates the shape of the numerical array that results from the computation performed by the array on its inputs. We indicate the zeroth dimension of these shapes as a user-definable batch size, but for the work in this paper this batch size is always one. The ``Parameter Count'' column lists the numbers of parameters (weights) that can be trained for each layer. Horizontal lines divide the table into separate blocks that are illustrated in Figure \ref{fig:discriminator_model}.}
    \label{tab:discriminator_model}
\begin{tabular}{lllr}
Layer & Type & Output Shape & Parameter Count \\
\hline
1 & InputLayer (Mould Image) & (Batch Size, 256, 256, 2) & 0 \\
2 & InputLayer (Label or Gen. Image) & (Batch Size, 256, 256, 8) & 0 \\
\hline
3 & Concatenate 1 and 2 & (Batch Size, 256, 256, 10) & 0 \\
\hline
4 & Conv2D & (Batch Size, 128, 128, 64) & 10240 \\
5 & LeakyReLU & (Batch Size, 128, 128, 64) & 0 \\
\hline
6 & Conv2D & (Batch Size, 64, 64, 128) & 131072 \\
7 & BatchNormalization & (Batch Size, 64, 64, 128) & 512 \\
8 & LeakyReLU & (Batch Size, 64, 64, 128) & 0 \\
\hline
9 & Conv2D & (Batch Size, 32, 32, 256) & 524288 \\
10 & BatchNormalization & (Batch Size, 32, 32, 256) & 1024 \\
11 & LeakyReLU & (Batch Size, 32, 32, 256) & 0 \\
\hline
12 & ZeroPadding2D & (Batch Size, 34, 34, 256) & 0 \\
13 & Conv2D & (Batch Size, 31, 31, 512) & 2097152 \\
14 & BatchNormalization & (Batch Size, 31, 31, 512) & 2048 \\
15 & LeakyReLU & (Batch Size, 31, 31, 512) & 0 \\
\hline
16 & ZeroPadding2D & (Batch Size, 33, 33, 512) & 0 \\
17 & Conv2D & (Batch Size, 30, 30, 1) & 8193 \\
\hline
\end{tabular}
\end{table}

\begin{figure}
    \centering
    \includegraphics[width=0.8\textwidth]{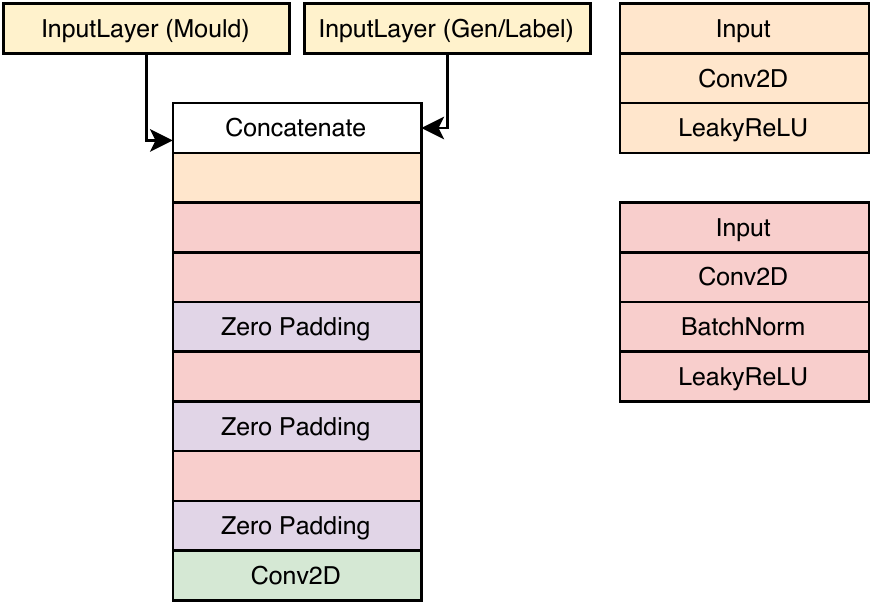}
    \caption{\textit{Left:} Schematic illustration of the \texttt{pix2pix} discriminator network. Similarly coloured blocks represent blocks containing layers that perform similar, but not necessarily identical, computational operations on their inputs. \textit{Right:} More detailed representations showing the layers that comprise each of the correspondingly coloured blocks in the network schematic. See Table \ref{tab:generator_model} for details of any differences between layers in different computational blocks. ``Concatenate'' blocks stack their inputs along the depth (channel) axis and output the result. Accordingly both inputs to a concatenate layer must have the same width and height, but their channel counts (depths) may be different.}
    \label{fig:discriminator_model}
\end{figure}

\end{document}